# Compressive Sensing and Morphology Singular Entropy-Based Real-time Secondary Voltage Control of Multi-area Power Systems

Irfan Khan, Vikram Bhattacharjee, Yinliang Xu, Soummya Kar, and Mo-Yuen Chow

*Abstract*-- This paper presents an improved secondary voltage control (SVC) methodology incorporating compressive sensing (CS) for a multi-area power system. SVC minimizes the voltage deviation of the load buses while CS deals with the problem of the limited bandwidth capacity of the communication channel by reducing the size of massive data output from phasor measurement unit (PMU) based monitoring system. The proposed strategy further incorporates the application of a Morphological Median Filter (MMF) to reduce noise from the output of the PMUs. To keep the control area secure and protected locally, Mathematical Singular Entropy (MSE) based fault identification approach is utilized for fast discovery of faults in the control area. Simulation results with 27-bus and 486-bus power systems show that CS can reduce the data size up to 1/10$^{th}$ while the MSE based fault identification technique can accurately distinguish between fault and steady state conditions.

## I. INTRODUCTION

Nowadays, distributed generation (DG) is increasing in the power system due to the environmental and economical motivation. The integration of massive DGs introduces intermittency, which makes the power system architecture susceptible to disturbances. These disturbances may cause severe voltage deviations or even voltage collapse in the power system. Therefore, adequate voltage control is very important to avoid voltage abnormalities in power systems. The hierarchical voltage control adopted in various countries can be considered as a solution to the problem of voltage collapse and instability [1]. Hierarchical voltage control usually consists of three control levels: primary, secondary and tertiary. The primary control is a local control and its main role is to automatically maintain the generator bus voltages close to their reference values by acting on generator exciters. However, primary voltage control is not always sufficient to cope with voltage violations associated with load changes. When the load increases, the primary control technique keeps the generator bus voltage close to its reference value. However, this control does not ensure the voltage regulation of all the load buses in the network.

To maintain the load bus voltage within the permissible limits, Secondary Voltage Control (SVC) is utilized in [2], which is a feedback control strategy to minimize the voltage deviation of load buses. SVC was first established in Europe and since then many researchers have utilized it in several countries like Brazil [3], Canada [4] and China [5]. For the implementation of SVC, the power system network is divided into several distinct zones where the voltage of each zone is controlled through automatic adjustment of voltage regulating units. The size of the adjustments is determined by the difference between the current load bus voltage and the predefined reference value of the bus. Thus, for the reliable functioning of SVC, all the load buses need to be monitored all the time. However, due to high cost, not all buses are equipped with PMUs for complete monitoring. The load buses which are equipped with PMUs, are called pilot buses while the rest are known as non-pilot buses [6].

Several SVC approaches have been proposed in the recent literature. Gu et al. [7] designed a non-linear state estimator based decentralized SVC scheme for voltage regulation which incorporates the effects of communication latency, whereas Wang et al. [8] designed a coordinated SVC strategy through a fuzzy logic scheme. As the number of pilot buses are usually less than the total number of load buses in a control area, the SVC problem can be mathematically viewed as an underdetermined problem. Several methods exist in the literature to solve the underdetermined SVC problem. Infinite norm regularization technique is used to minimize the worst-case voltage deviation [9]. Similarly, Chebyshev norm is utilized to design an optimal control strategy to minimize the worst-case voltage deviation.

The real-time monitoring of the load bus is usually accomplished using PMUs. Recent literature cites the usage of PMUs in power system applications [10], [11]. They are capable of providing fast and synchronized measurements and have the capability to overcome the interference from external conditions. The PMU measurements, however, may suffer from different issues related to cyber physical attacks, sensor failures and data congestion. The centralized SVC requires transmission of massive amount of data between the local control area and the central controller, which may cause data congestion. Data congestion may be significant, especially when the control area is large and the available bandwidth of the communication channel is limited. Bandwidth requirement is evaluated by [12], which states that a bandwidth of 5Mbps-10Mbps and 25–75 Mbps is required for applications within one control area and for inter control center communications, respectively. It may further be aggravated when data from several control zones is being concentrated at the Phasor Data Concentrator, before transmitting to the central controller. Thus, a data size reduction technique should be proposed to deal with these issues. Reduction or compression of data may also minimize the communication delay, the power loss on the communication channel, and communication channel cost.

CS is regarded as a promising joint data acquisition and reconstruction method to deal with the problem of limited bandwidth and data congestion. The data retrieved from the PMUs can be compressed before sending through the communication channel and can be recovered at the end of communication channel. Plenty of research has been performed to recover the signal at the end of communication channel [13], [14]. A CS based control strategy is developed for load frequency control in a multi-area power system network [15]. This strategy helps reduce the transmission data loss and increase the reliability of the communication network.

Performance of CS may be affected by the noise in the signal, which may induce error in the signal processing and degrade the system's dynamic performance [16]. In order to deal with such problems, several noise filtering techniques have been developed in the recent literature [17], [18]. Based on the theory of Mathematical Morphology (MM), a mathematical morphological filter (MMF) is proposed in [21], which helps smooth sharp edges, narrow valleys and gaps in a contour. The MMF filter has been found to be effective in the removal of signal noise in power systems applications. Apart from noise in the signals, abrupt load changes/faults in the power system may lead to significant deterioration of the terminal voltage. There has always been a need to design a fault identification technique that can observe changes in the signal waveforms locally and can activate the control action.

Recent literature in fault identification suggests MM based [19] and Morphology Singular Entropy (MSE) based [20] fault identification techniques. For rapid identification of disturbance, the signal should be analyzed before being sent to communication channels. Thus, a MSE-based fault identification technique is proposed which can detect the abnormality in the signal waveform accurately.

The contributions of this paper are summarized as follows:
1) Real-time SVC is proposed to minimize the worst-case voltage deviation of the control area by adopting an infinite norm minimization problem.
2) CS is used to encode and then decode the real-time power system data to deal with the issues of data congestion and limited bandwidth.
3) To deal with the noise in the signals, MMF is proposed to reduce the noise in the PMU output.
4) MSE based fault identification approach is employed for the fast identification of disturbances in the control area.

The prime control objective of the proposed control strategy is to improve the voltage profile of load buses in a multi-area power system which is achieved by using secondary voltage control. In addition to it, the communication burden of power system data transmission is decreased using CS. Another advantage of the proposed control scheme is to detect abnormal loading conditions/faults locally with the help of MSE.

To the best of authors' knowledge, no one has addressed all of these issues jointly before. Thus, contributions of this paper are a unified method that jointly addresses these problems.

The rest of this paper is organized as follows. Section II explains the CS, MMF and MSE, while Section III formulates the problem of SVC. The proposed CS and MSE based SVC is described in Section IV. Finally, results and conclusions are provided in Section V and Section VI, respectively.

## II. PRELIMINARIES

This section introduces the theoretical background of CS, MM and its MMF filter, Singular Value Decomposition (SVD) and the Shannon Entropy. It also relates MM, SVD and Shannon entropy to use them as a fault identification technique. In this paper, matrices are represented by bold and italic letters, vectors are represented by bold letters (not italic), the functions and variables in italic letters (not bold). The scalar quantities are neither bold nor italic.

### A. Compressive sensing

When transmitting power system state data over the restricted bandwidth communication channel, the amount and frequency of the data transmission have specific limits, such as Shannon's Channel Capacity, restrict the amount of digital information that can be transmitted over a given channel. That's why authors use CS based data exchange to make the data size smaller for fast data transmission.

There are a number of different compression algorithms, which can be broadly divided into two categories: lossless algorithms, and lossy algorithms. As the name states, lossless algorithms decrease the size of a given signal, while at the same time not losing any information from the original signal. For this reason, lossless compression algorithms are preferable to lossy algorithms, especially when the data needs to intactly arrive at the recipient. But lossless data compression algorithms may not guarantee compression for all input data sets. In other words, for any lossless data compression algorithm, there may be an input data set that does not get smaller when processed by the lossless algorithm. Also, lossless algorithms may not work perfectly for the dynamic nature of data, which is the case for power system output data. Thus, lossless compression may not be an ideal compression technique for this specific application. Another factor is that lossless compression cannot achieve a remarkable value of compression ratio whereas the objective in this paper is to compress the data size significantly.

On the other hand, using lossy compression algorithms, massive data can be compressed to a significant level. Very high compression ratio may be possible with lossy algorithms. But the quality loss of the retrieved data may be noticeable. There is a trade-off between compression ratio and quality of the recovered data. Data is quicker to be sent, transmitted and stored using lossy algorithms. Another advantage of lossy compression is the generic nature of its algorithms. Lossy compression algorithms may be applied to a wide variety of applications, especially to accommodate the dynamic nature of real-time data. As power system data is always changing, thus, lossy compression algorithms are the appropriate choice for this application. Therefore, based on these factors, it is recommended to use lossy algorithms for dynamic nature of power system data and CS is one of the most established compression technique for power system applications [15].

CS is a signal processing technique for efficiently acquiring and reconstructing a signal from a few measurements whose size is far smaller than that of the original signal. The theory of CS states that a sparse or compressible signal can be recovered with high probability from a few measurements, which is far smaller than the size of the original signal. Power system state signals, having spatial-temporal correlation, thus, can be regarded as compressible. Transmission of data using CS can be done by compressing the measurement data before sending it through communication channel and recovering it at the end of communication channel. Compression of measured data can be done using (1). If there are $n$ PMUs in the power system control area, at each time step $k$, the state measurements vector $\mathbf{x}(k)$ taken from $n$ sensors of the power system control area can be compressed into $m$ linear measurements (such that $m<n$, and $\mathbf{y} \in \mathbf{R}^m$) [21] as:

$$\mathbf{y}(k) = \boldsymbol{\Phi}\mathbf{x}(k) + \mathbf{N}(k) = \boldsymbol{\Phi}\boldsymbol{\Psi}\boldsymbol{\theta}(k) + \mathbf{N}(k) = \boldsymbol{\Upsilon}\boldsymbol{\theta}(k) + \mathbf{N}(k) \quad (1)$$

where $m$ is a function of the number of non-zero elements of the vector $\mathbf{x}(k)$, $\mathbf{x}(k) = \boldsymbol{\Psi}\boldsymbol{\theta}(k)$ is the representation of $\mathbf{x}(k)$ into its basis matrix $\boldsymbol{\Psi} \in R^{n \times n}$ and the vector of transform coefficients $\boldsymbol{\theta}(k) \in \mathbf{R}^n$. $\mathbf{N}(k)$ is the Gaussian additive white noise vector with zero mean and variance in the signal. $\boldsymbol{\Phi} \in R^{m \times n}$ is the measurement matrix and $\boldsymbol{\Upsilon} = \boldsymbol{\Phi}\boldsymbol{\Psi} \in R^{m \times n}$ is the sensing matrix. In this application measurement vector, $\mathbf{x}(k)$ is the voltage and reactive power magnitudes measured at the load buses. (1) will be used to compress measured data vector before sending it through communication channel.

At each time interval, each area of power system sends a vector of its states to Phasor Data Concatenator (PDC) which then concatenates all states into one big vector. For a window length of $w$, a matrix of dimension $n \times w$ at PDC can be obtained where $n$ is the number of states of whole power system. Then, CS is applied to compress the dataset.

For compression, by using DCT basis, the basis of the state measurements vector $\mathbf{x}(k)$ is first changed into transformation coefficients vector $\boldsymbol{\theta}(k)$. Then, the transformation coefficients vector $\boldsymbol{\theta}(k)$ is compressed into size $m \times 1$ using measurement matrix $\boldsymbol{\Phi}(k)$ of size $m \times n$ where $m<<n$. After compression, the compressed data is transmitted through communication channel. Note that, any universal data independent basis can be considered as a basis matrix $\boldsymbol{\Psi}$. Some of the available transformations may include the discrete cosine transform (DCT) basis, the discrete Fourier transform (DFT) basis, and the discrete wavelet transform (DWT), etc.

At the end of communication channel, recovering of the original measured data vector can be done by solving (1) for $\hat{\boldsymbol{\theta}}(k)$. However, decoding of (1) is an NP-hard problem involving $l_0$-norm minimization. Thus, $l_1$-norm minimization which is the convex approximation of $l_0$-norm is adopted:

$$\hat{\boldsymbol{\theta}}(k) = arg\,min_{\boldsymbol{\theta}(k)} \|\boldsymbol{\theta}(k)\|_1 \quad (2)$$
$$s.t. \quad \mathbf{y}(k) = \boldsymbol{\Upsilon}\boldsymbol{\theta}(k) + \mathbf{N}(k)$$

where $\hat{\boldsymbol{\theta}}(k)$ is the vector of transformation coefficients of the measurement vector. After recovering transformation coefficients vector $\hat{\boldsymbol{\theta}}(k)$, the state measurement vector is:

$$\hat{\mathbf{x}}(k) = \boldsymbol{\Psi}\hat{\boldsymbol{\theta}}(k) \quad (3)$$

where $\hat{\mathbf{x}}(k)$ is the vector of recovered measurements. Fig.1 shows the pictorial representation of CS where state measurement vector $\mathbf{x}(k)$ is being compressed using encoding algorithm of (1) and recovered using (2) & (3).

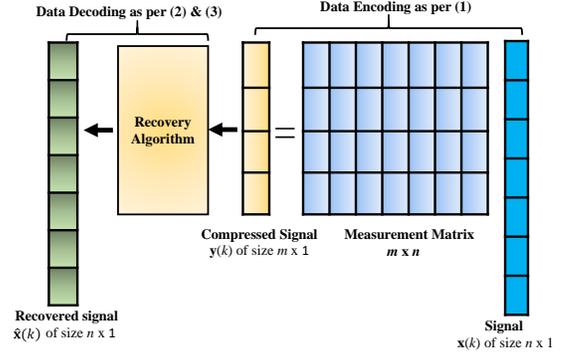

Fig. 1 CS based signal processing scheme

The performance of CS can be evaluated by the following two criterions:
1. Compression ratio ($\rho$):
$$\rho = \frac{\text{The size of data being transmitted without CS}}{\text{The size of data being transmitted with CS}} \quad (4)$$
2. Signal to error ratio (SNR):
$$SNR = -20(\log\|\mathbf{x} - \hat{\mathbf{x}}\|)/\|\mathbf{x}\| \quad (5)$$

In real-time system at each time instant, measured dataset for bus voltages are recorded using PMUs. However, in this paper, datasets are obtained by solving load flow at each time instant in the simulation studies.

### B. Mathematical Morphology

Mathematical Morphology (MM) is a concept based on the set theory [22] and has been established by introducing the fundamental operations on two sets. One of the set is the given measured signal and other set is a function known as the structuring element (SE) which is used to process the given signal. The SE is used as a probe which translates over the given signal.

The morphological operators are the set transformations which are highly efficient for feature extraction and de-noising the noisy signal while preserving the nature of the signal [23]. Two basic morphological operators are dilation and erosion which together make a pair of dual transforms. Let $\mathbf{x}(k)$ denote the state measurement obtained from power system and $g(\mathbf{s})$ denote the structuring element, then mathematically, the dilation ($\oplus$) and erosion ($\ominus$) of $\mathbf{x}(k)$ by $g(\mathbf{s})$ are given by:

$$(\mathbf{x} \oplus g)(k) = \max_{\mathbf{s}}(\mathbf{x}(k+\mathbf{s}) + g(\mathbf{s})) \quad (6)$$
$$(\mathbf{x} \ominus g)(k) = \min_{\mathbf{s}}(\mathbf{x}(k+\mathbf{s}) - g(\mathbf{s})) \quad (7)$$

where $\mathbf{s}$ is the vector of structuring element [24].

Dilation operator computes the maximum value within the neighborhood of $\mathbf{x}(k)$ and erosion operator calculates the minimum value within the neighborhood of $\mathbf{x}(k)$. Intuitively, dilation can be considered as swelling of $\mathbf{x}(k)$ while erosion can be imagined as shrinking of $\mathbf{x}(k)$. SE can have different shapes and sizes depending on the problem to be solved. A simple case of SE is of the form $g(\mathbf{s})=0$ and it is called (known as) "flat SE A structuring element is a shape that is used to probe an image. Flat structuring elements are binary structuring elements considered as masks. Most of the time, flat structuring elements are centered and are square matrices, with an odd size: 3x3, 5x5, etc. These matrices consist of 0's and 1's and are typically much smaller than the image being processed. It is common to use flat structuring elements in morphological applications. In order to filter the noise from the signal, MMF is designed by averaging the dilated and eroded waveforms as:

$$\mathbf{x}_{filter}(k) = \frac{1}{2}\left((\mathbf{x} \oplus g)(k) + (\mathbf{x} \ominus g)(k)\right) \tag{8}$$

where $\mathbf{x}_{filter}(k)$ is the filtered output of $\mathbf{x}(k)$. By applying different shapes and lengths of SE, filter can de-noise the input signal in different levels.

Length of window $w$ is an important factor to be decided which affects the speed and computational complexity of the control algorithm. Generally speaking, the longer the data window is, the longer the convergence speed will be and the heavier the computational burden will be. Therefore, the determination of the length of data window depends on the preference of the individual application for accuracy, response time, and computational burden. If an application has a preference about accuracy more than response time and computational burden, a long window length may be used. In this paper, a window length of 20 is used which is a reasonable window length for a good accuracy and convergence speed. Another important factor is the selection of the structuring element, the larger the scale of structuring element is, more details are removed from the input signal.

*C. Singular Value Decomposition (SVD)*

SVD is used to decompose a matrix $A \in R_r^{m_1 \times n_1}$ into two orthonormal matrices $L \in R^{m_1 \times m_1}$ and $M \in R^{n_1 \times n_1}$ and a rectangular diagonal matrix $\Sigma \in R^{m_1 \times n_1}$ as given in (9) [25]

$$A = L\Sigma M^T \tag{9}$$

Where $\Sigma$ consists of positive square roots of eigenvalues known as singular values of $A$.

Singular values ($\sigma$) are arranged in its descending order in $\Sigma$. By decomposing the matrix of measured state data using SVD, the state data can be represented in a decreasing order of their magnitudes using singular values.

*D. Information Entropy*

In information theory, information entropy is the expected value of the information contained in each message. In other words, it is the measure of the uncertainty in each data package as defined by (10) [26]

$$E = -\sum_{i=1}^{n_2} p_i \ln p_i \tag{10}$$

where $0 \le p_i \le 1$ is the probability of occurrence of an event, $\sum_{i=1}^{n_2} p_i = 1$ and $n_2$ is the total number of events or total number of data packages. When all probabilities are equal to each other, the information entropy $E$ reaches its maximum value $\ln(n_2)$ that means it is the most uncertain situation.

## III. SECONDARY VOLTAGE CONTROL

In the power system hierarchical control, secondary voltage control (SVC) deals with the regulation of load bus voltages only. It is the primary control/ local control that deals with the voltage regulation of source buses. As load buses have no sources connected on them, so primary control cannot regulate their states and thus, secondary control comes into operation and regulates the load buses. That's why this paper considers regulating the voltage of load buses in the proposed algorithm of secondary voltage control.

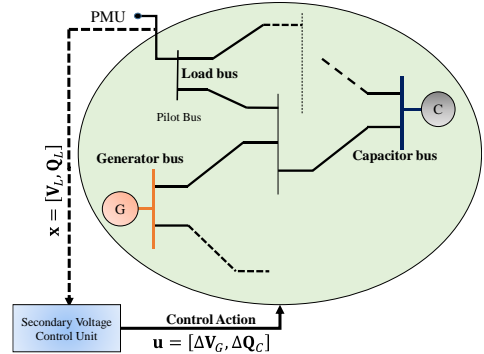

Fig. 2. Overview of the Secondary Voltage Control

SVC minimizes the load buses voltage deviations within the power system. Voltage deviation is calculated by the difference between the measured bus voltage and its predefined reference value as:

$$\Delta|\mathbf{V}_L| = |\mathbf{V}_L| - |\mathbf{V}_L^*| \tag{11}$$

where $\mathbf{V}_L$ is the load bus measured voltage and $\mathbf{V}_L^*$ is its predefined reference voltage. $\mathbf{V}_L$ is usually measured with the help of PMUs, attached on the selected buses as shown in Fig.2, which explains the working diagram of SVC in a power system network.

*A. Problem Formulation*

As discussed in Chapter I and II that in centralized control of SVC, the transmission of massive data incurs delays in data arrival at the receiving station. To deal with this problem in SVC, CS can be used to decrease the data size and hence decrease the communication delays in the data transfers. In CS based SVC, problem occurs when a fault or abnormal loading condition happens in power system. In this paper, we propose to minimize the worst-case scenario of the voltage deviation. As CS works on the assumption that temporal and spatial correlation exits in the power system data but in case of faults, this correlation may not exist. To deal with this problem, faults should be identified locally and alarm/protection devices should be activated to avoid any abnormal condition. Mathematical MSE has been used to address this problem in CS based SVC. Thus, the proposed control algorithm examines the voltage deviation of each bus in the control area and identifies the bus with the worst voltage deviation. The SVC algorithm generates the control output based on the maximum load bus voltage deviation and hence, it is performed by formulating it as an infinite norm minimization problem as:

$$\min_{\mathbf{u}} \|\Delta|\mathbf{V}_L|\|_\infty \tag{12}$$

where $\mathbf{u}$ is the control output of the SVC algorithm which includes the voltage regulating variables e.g. reactive power generation change from capacitor banks and voltage change of the voltage controlled buses etc. Solving the infinite norm minimization problem, generates the optimal **u** for which the maximum deviation is minimum.

The equality constraint of the SVC problem is given by power balance equations of the control area. Power balance equations are non-linear functions of the bus voltages (**v**), active power (**P**) injection and reactive power (**Q**) injection of the buses as given in (13). Non-linear equality constraints governing power flow in the control area of the power network have been described in detail in [10].

$$f(\mathbf{V}, \mathbf{u}, \mathbf{P}, \mathbf{Q}) = 0 \quad (13)$$

The inequality constraints of the optimization problem are governed by the upper and lower bounds of the control variable **u**. If the upper and lower bounds be denoted by $\bar{\mathbf{u}}$ and $\underline{\mathbf{u}}$ respectively, then the constraints are given by (14).

$$\underline{\mathbf{u}} \leq \mathbf{u} \leq \bar{\mathbf{u}} \quad (14)$$

*B. Proposed SVC approach*

DC power flow approximation has been recognized as valid approximation within real power P and phase angle. At the same time, DC approximation for reactive power and voltage magnitude V has also been explored by many recent researchers [27]-[29] and is considered to be a reasonable approximation for fast analysis of power system. The nonlinear equality constraint can be linearized using linearized model of decoupled power flow equations where real power changes are assumed to be less sensitive to changes in the voltage magnitude. Thus, the approximate model of the small disturbance voltage-VAR control can be represented by a linear relation between the reactive power and the voltages of the generator, capacitor and load buses as given in (15) [9, 10].

$$\begin{bmatrix} \Delta \mathbf{Q}_G \\ \Delta \mathbf{Q}_C \\ \Delta \mathbf{Q}_L \end{bmatrix} = \begin{bmatrix} \mathbf{B}_{GG} & \mathbf{B}_{GC} & \mathbf{B}_{GL} \\ \mathbf{B}_{CG} & \mathbf{B}_{CC} & \mathbf{B}_{CL} \\ \mathbf{B}_{LG} & \mathbf{B}_{LC} & \mathbf{B}_{LL} \end{bmatrix} \begin{bmatrix} \Delta |\mathbf{V}_G| \\ \Delta |\mathbf{V}_C| \\ \Delta |\mathbf{V}_L| \end{bmatrix} \quad (15)$$

where, ***B*** represents the system susceptance matrix, subscript *L* represents the load bus, subscript *G* represents the generator bus which can be used as voltage-controlled bus when $\Delta |\mathbf{V}_G|$ is considered as control variable. Subscript *C* represents the capacitor bank, which can also be used as voltage-controlled bus when $\Delta \mathbf{Q}_C$ represents the control variable.

From (15) the components $\Delta |\mathbf{V}_C|$ and $\Delta |\mathbf{V}_L|$ can expressed in (16) and (17) respectively.

$$\Delta |\mathbf{V}_C| = \mathbf{B}_{CC}^{-1} \Delta \mathbf{Q}_C - \mathbf{B}_{CC}^{-1} \mathbf{B}_{CG} \Delta |\mathbf{V}_G| - \mathbf{B}_{CC}^{-1} \mathbf{B}_{CL} \Delta |\mathbf{V}_L| \quad (16)$$

$$\Delta |\mathbf{V}_L| = \mathbf{B}_{LL}^{-1} \Delta \mathbf{Q}_L - \mathbf{B}_{LL}^{-1} \mathbf{B}_{LG} \Delta |\mathbf{V}_G| - \mathbf{B}_{LL}^{-1} \mathbf{B}_{LC} \Delta |\mathbf{V}_C| . \quad (17)$$

Substitute the term of $\Delta |\mathbf{V}_C|$ in (16) into (17) yields,

$$\Delta |\mathbf{V}_L| = \mathbf{B}_{LL}^{-1} \Delta \mathbf{Q}_L - \mathbf{B}_{LL}^{-1} \mathbf{B}_{LG} \Delta |\mathbf{V}_G| - \\ \mathbf{B}_{LL}^{-1} \mathbf{B}_{LC} [\mathbf{B}_{CC}^{-1} \Delta \mathbf{Q}_C - \mathbf{B}_{CC}^{-1} \mathbf{B}_{CG} \Delta |\mathbf{V}_G| - \mathbf{B}_{CC}^{-1} \mathbf{B}_{CL} \Delta |\mathbf{V}_L|] \quad (18)$$

From (18) it is evident that, the load bus voltage change $\Delta |\mathbf{V}_L|$ can be expressed as a function of the reactive power load change ($\Delta \mathbf{Q}_L$) and the control output (**u**), as given in (19).

$$\Delta |\mathbf{V}_L| = (\mathbf{J}_1 \Delta \mathbf{Q}_L - \mathbf{J}_2 \mathbf{u}) \quad (19)$$

where,
$$\mathbf{J}_1 = (\mathbf{B}_{LL} - \mathbf{B}_{LC} \mathbf{B}_{CC}^{-1} \mathbf{B}_{CL})^{-1} \quad (20)$$

$$\mathbf{J}_2 = \mathbf{J}_1 [\mathbf{B}_{LG} - \mathbf{B}_{LC} \mathbf{B}_{CG} \quad \mathbf{B}_{LC} \mathbf{B}_{CC}^{-1}] \quad (21)$$

$$\mathbf{u} = \begin{bmatrix} \Delta |\mathbf{V}_G| & \Delta \mathbf{Q}_C \end{bmatrix} . \quad (22)$$

By utilizing DC Decoupled power flow of (15), the equality constraint of (13) is already incorporated in (19). Thus, infinite norm minimization problem of (12) can be rewritten as:

$$\min_{\mathbf{u}} \| (\mathbf{J}_1 \Delta \mathbf{Q}_L - \mathbf{J}_2 \mathbf{u}) \|_\infty, \quad \text{s.t.} \quad \underline{\mathbf{u}} \leq \mathbf{u} \leq \bar{\mathbf{u}} . \quad (23)$$

The optimal solution of the (23) generates the optimal value of control input **u**, which consists of the vector of generator bus voltage change and the vector of capacitor bus reactive power generation change. Each reactive power source is limited with a maximum value $\bar{\mathbf{u}}$ and minimum value $\underline{\mathbf{u}}$. If the control input u exceeds its limits, the proposed algorithm will restrict it within its lower and upper bounds. Monitoring the vector of reactive power load change $\Delta \mathbf{Q}_L$, can be achieved by deploying PMUs on load buses. However, currently, all load buses are not equipped with PMUs and hence, it is required to design a method to estimate reactive power load changes on the non-pilot load buses.

For the case when partial load buses are equipped with PMUs, voltage measurements $\Delta |\mathbf{V}_p|$ are available only at the given pilot buses. It means $\Delta |\mathbf{V}_p|$ has only some elements of vector $\mathbf{J}_1 \Delta \mathbf{Q}_L$. Using only the rows of corresponding pilot buses in $J_1$ and denoting it as $J_p$, the optimal reactive power load change on all buses $\Delta \mathbf{Q}^*_L$, can be estimated by solving the following underdetermined system [9] as:

$$\Delta \mathbf{Q}^*_L = \mathbf{J}_p^T (\mathbf{J}_p \mathbf{J}_p^T)^{-1} \Delta |\mathbf{V}_p| . \quad (24)$$

After obtaining $\Delta \mathbf{Q}^*_L$, infinite norm minimization problem in (23) can be solved for the optimal **u**. Based on the new control input $\mathbf{u}^{k+1}$ for every new iteration, $\Delta |\mathbf{V}_L^{k+1}|$ is calculated using (19), as shown in Fig. 3. The Pilot bus voltage deviation, $\Delta |\mathbf{V}_p^{k+1}|$ can be taken out from the set consisting of the deviation of all the load buses to calculate the pilot bus voltage $|\mathbf{V}_p^{k+1}|$ for the next iteration as given in (25).

$$|\mathbf{V}_p^{k+1}| = |\mathbf{V}_p^k| + \beta \Delta |\mathbf{V}_p^{k+1}| \quad (25)$$

where, an acceleration parameter $\beta$ with a value of 0.5, is incorporated in (25) to further speed up the update process.

The proposed algorithm provides an optimal control output when the predefined value of tolerance $\varepsilon$ is reached:

$$||\mathbf{V}_p^{k+1}| - |\mathbf{V}_p^k|| < \varepsilon \quad (26)$$

where $\varepsilon$ is the predefined value of tolerance, which is chosen as 0.001 in this paper.

Once the predefined value of tolerance is achieved, the proposed algorithm will stop and produce the optimal **u***.

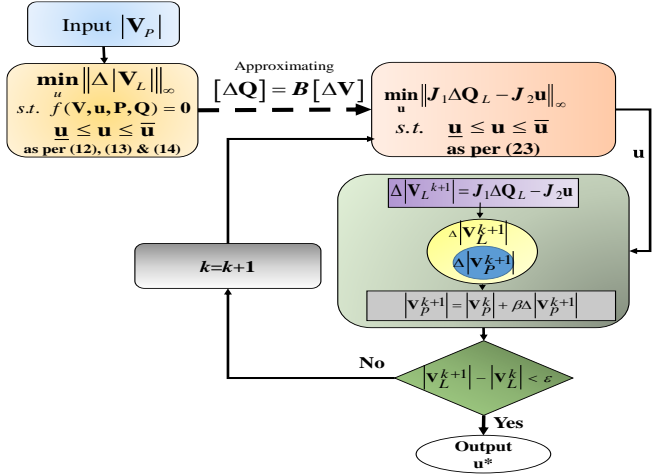

Fig. 3. Flowchart for the SVC algorithm

The performance of the SVC algorithm can be expressed as the root mean square value of voltage deviations at all load buses given by (27)

$$x_{rms} = \sqrt{\frac{1}{l}\sum_{i=1}^{l}(|\mathbf{V}_L|-|\mathbf{V}_L^*|)^2} \qquad (27)$$

where $l$ is the total number of load buses.

## IV. PROPOSED CS AND MSE BASED SVC ALGORITHM

SVC of the power system is performed by dividing it into several control areas as shown in Fig. 4. In this paper, pre-recorded data is not used rather the real-time data of $\hat{\mathbf{x}}(k) = \begin{bmatrix} \hat{\mathbf{V}}(k) & \hat{\mathbf{Q}}(k) \end{bmatrix}$ is measured at each instance $k$. Here, real time power system states data from $i$th control area $\mathbf{x}_i(k)$ consist of load bus voltage and load bus reactive power load demand are measured online at each instance. It is assumed that data measuring techniques ideal and measured data has no uncertainties. These measured states may introduce some random noise due to sensor inefficiency or harmonics present in the power system. To deal with such scenarios, $\mathbf{x}_i(k)$ is passed through MMF to eradicate the noise from the signal. Before sending the system data to the central controller for system state observation and generation of a control output, it is recommended to check the system states locally for any fault or abnormal condition. For fast local control action for faults, MSE is calculated locally and a control signal is sent back to the control area to raise an alarm or initialize the relay when value of entropy exceeds a certain pre-set limit. Following steps are taken to calculate the MSE:

1) Suppose a signal with a data window length of $w$ contains $w \geq 1$ consecutive readings of a signal $\mathbf{x}$ at the time step $k$ is given as $\mathbf{x}(k) = \begin{bmatrix} x(k-w+1) & \cdots & x(k) \end{bmatrix}$. Now, MMF defined in (8) is applied to construct a matrix $H$ from $\mathbf{x}(k)$. By choosing a flat SE of different sizes with its origin in the middle, matrix $H$ is given as

$$H = \begin{bmatrix} h_1(k-w+1) & \cdots & h_1(k) \\ \vdots & \ddots & \vdots \\ h_{m_2}(k-w+1) & \cdots & h_{m_2}(k) \end{bmatrix}_{m_2 \times w} \qquad (28)$$

where $h_i(k)$ is the output of MMF filter at $\mathbf{x}(k)$ for different length of SE when length of SE is increased from 1 to $m_2$ at a step of 1. First row of $H$ is the same as the input data $\mathbf{x}(k)$. As the length of SE increases, more details are eliminated from $\mathbf{x}(k)$ [20] which implies that different rows preserve information at different levels.

2) After formation of $H$, it is decomposed using *SVD* and only those singular values can be selected which represent the main components of the signal.
3) Assume, $m_3$ is the number of selected singular values in step 2, then the probability $p_j$ associated with $\sigma_j$ is

$$p_j = \sigma_j \Big/ \sum_{j=1}^{m_3} \sigma_j \qquad (29)$$

4) Finally, entropy ($E$) of $\mathbf{x}(k)$ can be calculated as per (10).

It can be further explained from (10) and (29) that $E$ will be zero for a pure sinusoidal input $\mathbf{x}(k)$ because in this situation, only first singular value is nonzero with a probability of 1. Any change in the signal will lead to a higher entropy.

Data from each control area is concatenated at PDC and sent to the central control through communication channel. To reduce the data size, CS is utilized to compress the data and recover it at the end of communication channel. In the application of SVC, the recovered data is given as (21)

$$\hat{\mathbf{x}}(k) = \begin{bmatrix} \hat{\mathbf{V}}(k) & \hat{\mathbf{Q}}(k) \end{bmatrix} \qquad (30)$$

where $\hat{\mathbf{V}}(k)$ and $\hat{\mathbf{Q}}(k)$ are the vector of voltage and reactive power in power system, respectively.

The data obtained from monitoring a physical system e.g. power system, inherently contains spatial and temporal correlation. Because of this inherent property, the monitored data can be compressed before transmitting through the communication channel. Likewise, it can be recovered successfully at the end of communication channel using various recovery techniques. In this paper, the Orthogonal Matching Pursuit (OMP) algorithm is utilized to solve the optimization problem. OMP is a recursive algorithm and provides faster rate of convergence especially in the case of non-orthogonal directories [29].

Finally, the central controller decides the SVC control action based on the recovered data and sends it back to each control area as explained in Fig. 4.

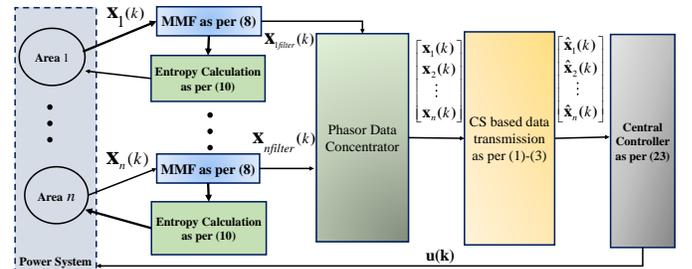

Fig. 4. Extensive working of the proposes CS and MSE base SVC algorithm

## V. SIMULATION RESULTS & DISCUSSION

This chapter shows the simulation results for two power systems: 27-bus power system and 846-power system. Both power systems are further divided into three sub-areas.

## A. Case Study 1: 27-bus Power System

The proposed CS & MSE based centralized SVC algorithm is applied to the 27-bus power system as shown in Fig 5. It is built by connecting 3 control areas with modified IEEE 9- bus system [30] in each area. The impedances for all the tie-lines are set to 0.02+j0.07 p.u. In each area, bus no. 7, 8 and 9 are selected as load buses. In this case, all the load buses are assumed to have PMU on them for monitoring of voltage deviation. However, results are shown in the section B of this chapter about using different number of pilot buses in this power system. Bus no. 2 and 3 are equipped with capacitor banks while 4, 5 and 6 are voltage controlled buses, which can change their bus voltage to improve the voltage profile of the load buses. To test the proposed algorithm in real-time conditions, a series of load changing events are introduced on various buses as given in Table. I.

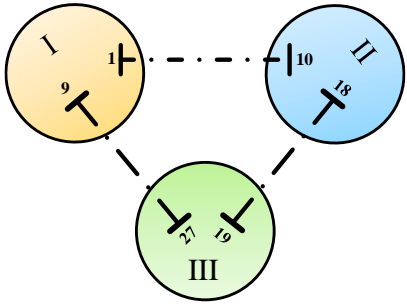

Fig. 5. 27-bus power system, comprising of three control areas of 9 buses in each control area

TABLE.I EVENT SEQUENCES OF LOAD CHANGES ON 9-BUS CONTROL AREA

| Time stamp | Event No. | Bus No | Load Type | Load Change |
|---|---|---|---|---|
| 21s | Event1 | 7,8,9 | Reactive Load | 1.03×Initial |
| 42s | Event2 | 7,8,9 | Reactive Load | 0.97×Event1 |

For the results in Figs. 6-10, compression ratio of 2 is used. Various values of compression ratios are tested to investigate its effect on the system performance and results are shown in the tabular form in part C of this Section. The voltage profile improvement of the load buses in control area I is shown in Fig. 6. It shows that SVC algorithm is very fast in improving the load bus voltage close to the reference value. To analyze the behavior of the algorithm in case of changing reactive power loads, load voltages falls down at 21s when the reactive power load is increased by 1.03 times and voltage improves at 42s when reactive power load decreases by 0.97 times. The response of the proposed control algorithm during instances of load changes substantiates its performance and adaptability. RMS value of the voltage deviation is calculated using (27) and shown in Fig. 7. It is clear that proposed algorithm minimizes the voltage deviation of the load buses, where overloaded buses have higher voltage deviations. The optimal control input which minimizes the voltage deviation in control area I of the system is shown in Fig. 8. It clearly exhibits that it converges to its optimal value to regulate the voltage deviation in the system.

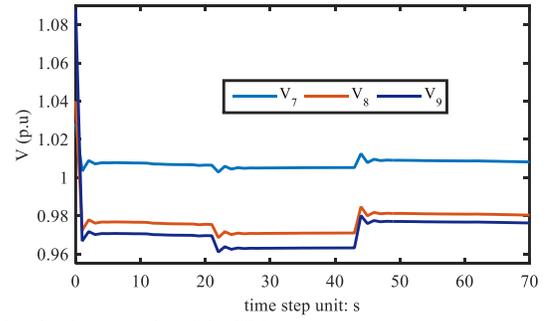

Fig. 6. Load voltage results of the 27-bus power system

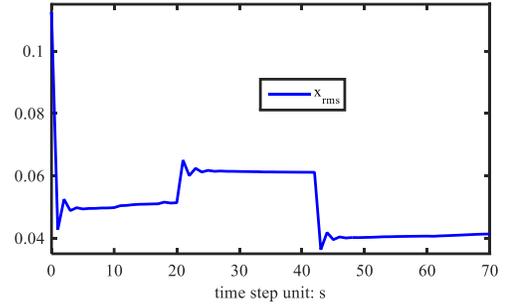

Fig. 7. $X_{rms}$ value of voltage deviation of the 27-bus power system

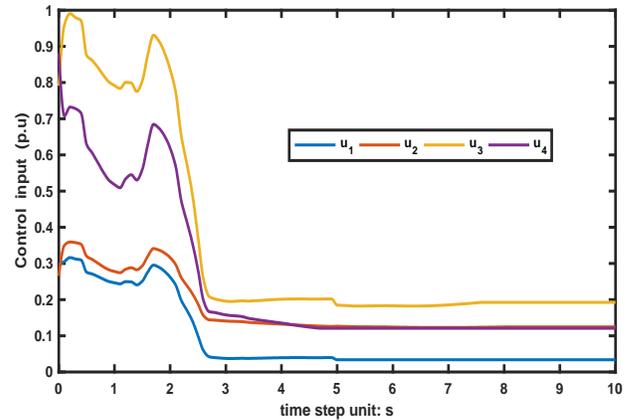

Fig. 8. Optimal control input for SVC in control area I

To test the effectiveness of MMF, white Gaussian noise with zero mean and 40 SNR is introduced in the load bus voltage of bus 8 at 30s as shown in Fig. 8. It shows that noise in the voltage signal of bus 8 also introduces small noise in its neighboring buses. Thus, it is important to use the filtered signal for control of the system.

From Fig. 9, it can be observed that the proposed MMF reduces the noise peaks in the output voltage signal. To investigate the working of fault identification by calculating its MSE, the MSE of the noisy voltage signals of bus 7-9 is calculated and shown in Fig. 10. It exhibits that the maximum value of MSE due to noise is for the voltage signal of bus 8 and its value oscillates between 0 and 0.015 nat. This value is far less than 0.05 which is the pre-set value, to send an actuation signal to the protection relay. Thus, in this case the proposed control strategy will assume it as noise and will not send a triggering signal to the relay. To test the working of proposed MSE method for fault identification, reactive power load is increased by 1.75 times on bus 8 at $20^{th}$ second as shown in Fig. 11 where X represents the X-axis value of time step in seconds while Y represents the Y-axis value of voltage in p.u. and entropy in nat. Due to this severe overloading condition, $V_8$ falls to 0.8454 p.u. and the measured entropy rises to 0.08685 nat. This value is higher than the pre-set threshold value of 0.05 and thus an actuation signal will be sent to the protection relay to activate the circuit breakers or raise an alarm.

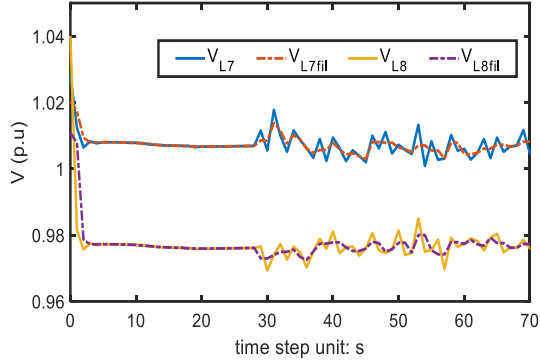

Fig. 9. Voltage profile with & without MMF

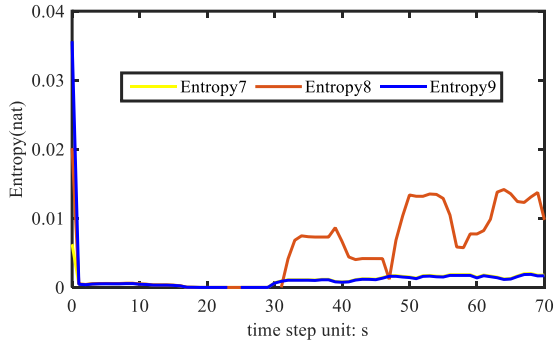

Fig. 10. Entropy variation of voltage profiles bus 7, bus 8 and bus 9

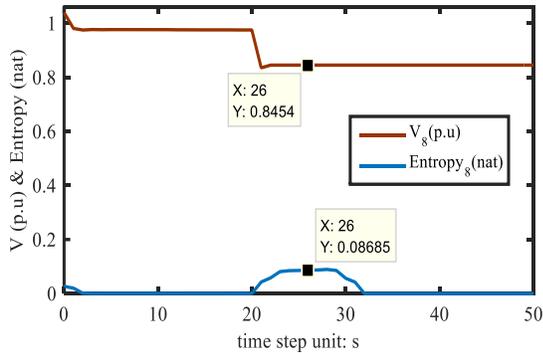

Fig. 11. Entropy change when $V_8$ falls to 0.85 p.u.

The proposed SVC algorithm will work satisfactory as long as at least one load is a pilot bus. To find the optimal control input using (23), we need to calculate $\Delta Q_L$ using (24) and $\Delta Q_L$ will be undefined if $J_p$ is zero or in other words, no load bus is selected as pilot bus. Thus, at least one of the load bus need to be attached with PMU. For 27-bus power system, authors have performed simulations for different number of pilot buses. Total number of load buses in 27-bus power system are 9. Authors varied pilot buses from one to nine and calculated the root mean square value of voltage deviation for each case. Fig.12 shows how increasing number of pilot buses improves the voltage deviation. That's why it is always recommended to connect a PMU on each load bus of power system.

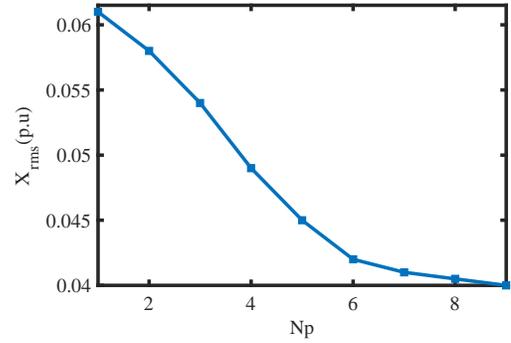

Fig.12. Number of pilot buses vs root mean square value of voltage deviation

### B. Case Study 2: 486-Bus System

To test the proposed algorithm's performance on a large-scale control area, a test case on 486-bus power system, built by connecting 3 areas with 162-buses in each area as shown in Fig 13, is conducted here. Parameters of the 162-bus system can be found in [30], where each control area contains 32 generators and 50 capacitor banks to regulate the voltage of load buses. The impedances for all the tie-lines are set to 0.028+j0.096 p.u. To test the proposed algorithm on the real-time power system, reactive power loads of the control area III are changed to 1.20 times and 0.80 times of the initial loads on all load buses at $t$=10s and $t$=20s, respectively.

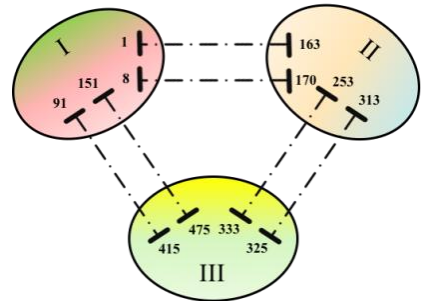

Fig. 13. 486-bus power system, comprising of three control areas of 162 buses in each control area

Voltage profile of the load buses in control area III is shown in Fig. 14. It exhibits that the voltage profile improves from 1.15 p.u. to a permissible operating voltage range of 0.95 p.u. to 1.05 p.u. As the reactive power load increases at $t$=10s, voltage profile deteriorates. However, the proposed algorithm still keeps it within the permissible operating range. Similar is the case when reactive power load is dropped at $t$=20s. RMS value of the voltage deviation of the control area III is shown in Fig. 15, which shows that load change increases the voltage deviation in the system.

To analyze the effect of CS on the communication delay of data transfer, proposed SVC algorithm is implemented without using CS. Real time data of the three control areas is transferred through the communication channel simultaneously. The obtained results are then compared with the results obtained by using CS for data reduction as shown in Fig. 16. The solid lines show plots without CS and dotted lines show plots with CS. When CS is utilized to compress the power system data, it decreases the communication delays during the data transfer. It is clear that the use of CS relieves the communication congestion and thus, decreases the communication delay during data transfer.

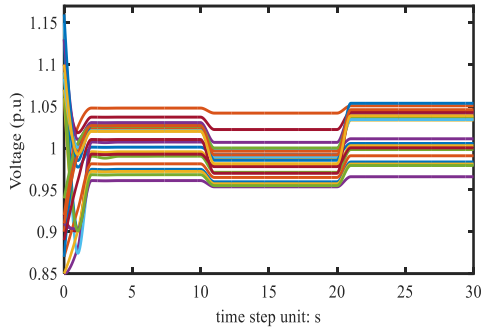
Fig. 14. Load bus voltage results of the 486-bus system in control area III

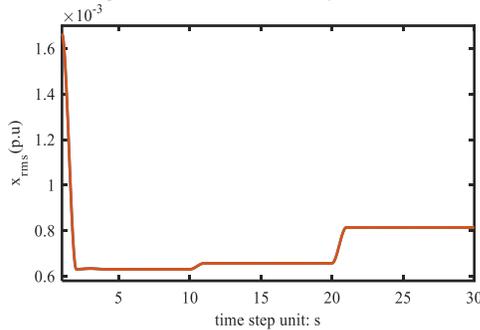
Fig. 15. RMS of the load bus voltage deviation of the 486-bus system in control area III

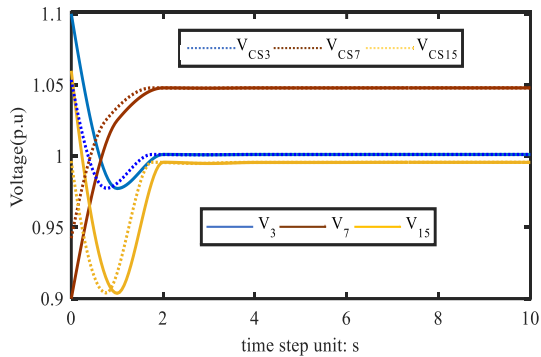
Fig. 16. Voltage profile comparisons under the communication delay

### C. Achievable Compression Ratio

CS based SVC algorithm is tested for various compression ratios and its corresponding SNR of the recovered signal. It is observed that as the compression ratio of the CS increases, the quality of the recovered signal deteriorates, as shown in Table II. It shows that the maximum compression ratio achieved for 27-bus power system is 6, after which the recovered signal does not produce any useful state information. The maximum compression obtained for 486-bus power system is 10. It implies that as the size of power system increases, the achievable compression ratio grows. It means use of CS is more effective for large power systems where data size is massive. Thus, the results corroborate that the proposed CS based SVC algorithm can reduce data congestion significantly.

TABLE II. VALUES OF ACHIEVABLE COMPRESSION RATIOS

| 27- bus power system | | 486-bus power system | |
|---|---|---|---|
| Compression Ratios ($\rho$) | SNR | Compression Ratio ($\rho$) | SNR |
| 1.5 | 86.76 | 2.0 | 90.67 |
| 3.0 | 75.14 | 4.0 | 81.39 |
| 4.0 | 66.30 | 6.0 | 68.01 |
| 5.0 | 53.49 | 8.0 | 52.82 |
| 6.0 | 36.87 | 10.0 | 35.65 |
| 7.0 | - | 12.0 | - |

### D. Discussion of primary vs. fast secondary control

Secondary voltage control is observed to be very fast and can converge to its optimal value within couple of seconds. However, this fast SVC cannot disturb the primary control in source buses because operating time of secondary is far less than the primary control. Primary control usually operates in micro/ milliseconds whereas the proposed fast secondary control takes few seconds to change the control variable settings. In real-time scenarios, secondary control is a centralized control that observes states of individual area, concatenates states of multiple areas, compresses them, transmits them through unreliable communication channel, recovers the original states at the end of communication channel and sends to the central controller that calculates the control variables and sends back to the system through communication channel. These all activities consume time which is much higher than the primary control that is a local control with no communications between local power system and the central controller.

## VI. CONCLUSION

This paper incorporates a CS based technique to reduce the data size for meeting the bandwidth requirement for real-time secondary voltage control of large power system networks. The obtained results demonstrate that CS decreases the communication delay during the data transmission for large power systems by reducing the data size to a factor of 10X. Further, the MMF is introduced to eradicate the noise present in the states of the power system. Furthermore, MSE based fault identification technique is proposed to clear faults locally in the control area in a fast and efficient way. Results show that in case of faults or abnormal conditions, the entropy of the RMS value of the voltage exceeds the preset limit and a switching signal is sent to the relay to raise an alarm or activate the breakers.